\documentclass[aps,prl,twocolumn]{revtex4-1}

\usepackage{mathtools}
\usepackage{amssymb}
\usepackage{graphicx}
\usepackage{bbold}
\usepackage{hyperref}
\usepackage{xcolor}
\usepackage{enumerate}
\usepackage{xspace}
\usepackage[normalem]{ulem}

\hypersetup{colorlinks=true, linkcolor=blue, urlcolor=blue, citecolor=blue}

\begin{document}
\title{Correlated quantum machines beyond the standard second law}

\author{Milton Aguilar}
	\affiliation{Institute for Theoretical Physics I, University of Stuttgart, D-70550 Stuttgart, Germany}

\author{Eric Lutz}
	\affiliation{Institute for Theoretical Physics I, University of Stuttgart, D-70550 Stuttgart, Germany}

\begin{abstract}
 The laws of thermodynamics strongly restrict the performance of thermal machines. Standard thermodynamics, initially developed for uncorrelated macroscopic systems, does not hold for microscopic systems  correlated with their environments. We here  derive an exact formula for the efficiency of \textit{any} cyclically driven quantum engine by using generalized laws of quantum thermodynamics that account for all possible correlations between all involved parties, including initial correlations. Furthermore, we demonstrate the existence of two basic modes of engine operation: the usual thermal case, where heat is converted into work, and a novel athermal regime, where work is extracted from entropic resources, such as system-bath correlations. In the latter regime, the efficiency is not bounded by the usual Carnot formula. Our results provide a unified formalism to determine the  efficiency of correlated microscopic quantum machines.
\end{abstract}

\maketitle

\noindent \textbf{INTRODUCTION}\\
\noindent Thermodynamics offers a powerful framework to describe the equilibrium properties of macroscopic systems. By providing quantitative relationships between observable quantities, it allows one to predict the state of arbitrary systems when external parameters are varied \cite{pip66}. An important application of the formalism is the study of the interconversion of different energy forms, such as mechanical, chemical and thermal energies. In particular, the  laws of thermodynamics restrict the efficiency of heat-work conversion in cyclic processes \cite{pip66}. The maximum achievable value of the efficiency,  defined as the ratio of work output and heat input, of any heat engine  coupled to two thermal baths is thus given by the Carnot formula, $\eta_\text{C} = 1-T_\text{c}/T_\text{h}$, where $T_\text{c,h}$ are  the respective temperatures of the  cold and hot heat reservoirs \cite{pip66}.

Despite its successes, standard thermodynamics relies on the fundamental hypothesis that system and baths are uncorrelated \cite{lan76}. This assumption is well justified for large classical systems that weakly interact with their reservoirs, since system-bath coupling energies, which typically scale like the surface of the systems, are usually much smaller than their internal energies, which scale like their volume \cite{lan76}. However, this condition is often violated for microscopic quantum systems owing to the presence of strong interactions \cite{tal20} and/or quantum correlations \cite{lan21} between system and reservoirs. In particular, system and baths can be entangled, even for weak coupling, below a critical temperature \cite{eis02,hil09,wil11,per11,ank14,ros18}. As a consequence, the usual laws of macroscopic thermodynamics no longer apply in this regime -- and need to be generalized \cite{all00,car16,ber17,per18,str19,mic19,lip24,sap19,riv20,cre21,liu21}. While  several methods to include correlations between a system and one or two reservoirs have been put forward \cite{all00,car16,ber17,per18,str19,mic19,lip24,sap19,riv20,cre21,liu21}, including an information-theoretic framework \cite{ber17}, a repeated interaction scheme \cite{str19} or a 'Hamiltonian of mean force' approach \cite{riv20,cre21}, a general formalism that allows one to explicitly account for all possible quantum correlations, within the system, between the system and many baths, as well as between the baths is missing.

Quantum machines that convert one form of energy into another, such as  engines and refrigerators \cite{kos13,kos14}, aim at harnessing quantum effects to improve their performance  \cite{ben17,mye22,arr23,can24}. Quantum phenomena, including quantum coherence, quantum correlations and squeezing, may indeed strongly impact thermodynamic processes, and have the potential to be exploited as a  resource \cite{ben17,mye22,arr23,can24}. Initial correlations have, for example, been found to  induce heat flow reversals \cite{mic19,lip24}, highlighting the need to generalize the second law in this situation \cite{par08,jen10}.  In addition, coherence \cite{scu03,scu11}, correlations \cite{dil09,per15}, as well as squeezing \cite{ros14,nie18}, have been shown to increase the efficiency of cyclic machines, and even surpass the Carnot bound in  some instances. However, a unified framework allowing one to describe the performance of quantum correlated machines is currently not available. Especially, their maximum  possible efficiency in such broad setting is currently unknown.

We here formulate universally applicable extensions of the first and second laws  for a generic quantum system subjected to periodic driving and coupled to an arbitrary number of reservoirs via general, not necessarily energy conserving, interactions. We obtain an \textit{exact}, generalized Clausius equality that  accounts for all possible quantum correlations at all times. We uncover a new nonequilibrium contribution to the entropy balance that, unlike the entropy production, can be negative. We show that a negative sign is associated with a novel  operation regime in which   engines  extract work out of entropic resources, such as system-bath correlations,  instead of heat. We moreover derive a generalized formula for their maximum efficiency  that  contains this new entropic contribution, and demonstrate that it may exceed the standard Carnot limit. Finally, we illustrate our results by analyzing the performance of a two-oscillator quantum engine \cite{k84,qua06,ajm08}.\\

\noindent \textbf{RESULTS}\\
\noindent\textbf{Generalized laws of quantum thermodynamics} \\
 We consider a generic  quantum system $\cal S$ that interacts with an arbitrary number of reservoirs ${\cal R}_j$, with Hamiltonians $H_j$. We assume that the initial state $\rho_j(0)$ of each bath can be assigned a temperature $T_j$, such that  its mean energy is that of a thermal state $\rho_j^\text{th}$ at the same temperature, $\text{tr} [ \rho_{j} (0) H_{j}] = \text{tr} [ \rho_{j}^\text{th} H_{j}]$; this is  the most common approach to assign a temperature to reservoirs that are not necessarily in  equilibrium \cite{lbplb2023}. In order to account for correlations within the system, we divide it into a collection of noninteracting subsystems $\mathcal{S}_{i}$ with Hamiltonians $H_i$; each  subsystem may include many interacting parts. In the sequel, we will use the indices $i$ ($j$) to  refer to quantities  related to the subsystems  (baths). The total Hamiltonian is then  $H_\text{tot} (t) = \sum_{i} H_{i} (t) + \sum_{j} H_{j} + \sum_{i,j} H_{ij} (t)$, where $H_{ij} (t)$ is the time-dependent interaction between subsystem $i$ and bath $j$. We stress that we do not take it to be  energy preserving \cite{cic22}, implying that we are not limited to the weak-coupling regime. We  additionally focus on cyclic processes with periodic system Hamiltonians, $H_{i} (t) = H_{i} (t + \tau)$, with period  $\tau$.

We begin by writing down  a generalized first law of thermodynamics. To that end, we employ standard definitions of work and heat valid for generic system-bath couplings \cite{esp10,stra17}. The total work extracted during a cycle is evaluated as $W =\int_{t}^{t + \tau} dt^{\prime} \langle \partial_{t^{\prime}} H_\text{tot} (t^{\prime}) \rangle$ \cite{liu21}. Since  the dynamics of the total system is unitary, the
energy change is solely due to the time dependence in the
Hamiltonian, which may thus be identified as work \cite{pus78}. This
approach naturally avoids the problem of the partitioning of the system-bath interaction \cite{bru16,new17,wie20,can21}. Using the Ehrenfest theorem for $H_\text{tot}(t)$ \cite{aul09}, one obtains the exact energy balance (Supplementary Materials)
\begin{equation}
	W = \sum_{j} Q_{j} + \Delta U,
	\label{1}
\end{equation}
where  $Q_{j} = \langle H_{j} \rangle (t + \tau) - \langle H_{j} \rangle (t) $ is the heat absorbed by reservoir $\mathcal{R}_{j}$, and $\Delta U = \sum_{i} \Delta U_{i} = \sum_{i} [ \langle H_{i} + \sum_{j} H_{ij} \rangle (t + \tau) - \langle H_{i} + \sum_{j} H_{ij} \rangle (t) ]$ is  the change of the energy of the compound system, including  the interaction energy. Expression \eqref{1} thus contains all the different forms of energy exchange, including the contribution originating  from the modulation of the system-bath coupling. The standard first law for   cycles \cite{pip66}, $W_{\mathcal{S}} = \sum_{j} Q_{j}$, which relates the work produced by the system, $W_{\mathcal{S}} =\sum_{i} \int_{t}^{t + \tau} dt^{\prime} \langle \partial_{t^{\prime}} H_{i} (t^{\prime}) \rangle$, to the total heat  is recovered when the internal energy remains constant over one period, and the system-bath coupling is energy conserving on average, $\sum_{i, j} \int_{t}^{t + \tau} dt^{\prime} \langle [ H_{i j } (t^{\prime}) , H_{i} (t^{\prime}) + H_{j} ] \rangle = 0$ \cite{cic22}. Note that a periodic Hamiltonian, $H_{i} (t) = H_{i} (t + \tau)$, does not necessarily imply a periodic averaged Hamiltonian, $\langle H_{i} \rangle(t) \neq \langle H_{i}\rangle (t + \tau)$, meaning  that in general $\Delta U \neq 0$ over one cycle. This is, for example, the case for systems that are coupled to finite reservoirs for which the state $\rho_i(t)$ of subsystem $i$ is not periodic \cite{ond81,wan16,taj17,poz18,moh19,ma20}.

A generalized  second law   can be similarly obtained  by computing the change of the von Neumann entropy, $S(\rho)=- \text{tr}[ \rho \ln (\rho)]$, for each subsystem and each bath over one cycle. Using the unitarity of the total time evolution, we have  $\sum_{i} \Delta S (\rho_{i})+ \sum_{j} \Delta S (\rho_{j}) = \Delta I (\mathcal{S} , \mathcal{R}) + \Delta C (\mathcal{S}) + \Delta C (\mathcal{R})$, where $I (\mathcal{S} , \mathcal{R}) = S (\rho_{\mathcal{S}}) + S (\rho_{\mathcal{R}}) - S (\rho_{\mathcal{SR}})$ is the mutual information between the system $\mathcal{S}$ and the collection $\mathcal{R}$ of all reservoirs \cite{cov91}.
 {The total correlations,  $C (\mathcal{S}) = \sum_{i} S (\rho_{i}) - S (\rho_{\mathcal{S}})$ and $C (\mathcal{R}) = \sum_{j} S (\rho_{j}) - S (\rho_{\mathcal{R}})$,   are moreover multivariate extensions of the mutual information \cite{w1960,mod12} that quantify  correlations among all the subsystems and among  all the baths.} The above equality can be understood as a conservation law relating the entropies of each party ($S (\rho_{i})$, $S (\rho_{j})$) and the correlations ($I (\mathcal{S} , \mathcal{R})$, $C (\mathcal{S}) $, $C (\mathcal{R}) $) between them. It can be rewritten in the form  (Supplementary Materials)
\begin{equation}
	\sum_{i}  \Delta S (\rho_i) + \sum_{j} \frac{Q_{j}}{kT_{j}} = \Delta \Sigma,
	\label{2}
\end{equation}
 where  $\Delta \Sigma = \Sigma (t + \tau) - \Sigma (t)$ with 
$\Sigma  = I (\mathcal{S} , \mathcal{R}) + C (\mathcal{S}) + C (\mathcal{R}) + \sum_{j} D(\rho_{j} || \rho_{j}^\text{th})$ ($k$ denotes the Boltzmann constant). The last  sum  contains the relative entropy, $D(\rho_j || \rho_j^\text{th}) = \text{Tr} \rho_j \ln \rho_j - \text{Tr} \rho_j \ln \rho_j^\text{th}$ \cite{cov91}, between the   reduced state $\rho_j$ of  reservoir $\mathcal{R}_{j}$ and the associated  reference thermal  state, $\rho_{j}^\text{th} = \exp ( - H_{j} / k T_{j}) / Z_{j}$ with partition function $Z_j$. The term $\Sigma(t+\tau)$ is the nonequilibrium entropy produced during the cycle of duration $\tau$ \cite{lan21}, here expressed as the amount of correlations created between system and baths, as well as within the system and  among the baths. It is  nonnegative and hence quantifies  the irreversibility of the  process \cite{lan21}. 
	
The general form \eqref{2} of the entropy balance  takes into account all possible sources of irreversibility, including  the noncyclic behavior of the system, the displacement out of local thermal equilibrium of the reservoirs, and the creation (or destruction) of classical and quantum correlations between all  parties. It thus extends previous microscopic generalizations of the second law that were obtained in the absence of initial correlations  \cite{esp10,stra17}. It contains, in particular, a new entropic contribution $\Sigma(t)$ related to the presence of  correlations at the beginning of a cycle. Whereas $\Sigma \geq 0$, the difference $\Delta \Sigma$ has no definite sign: $\Delta \Sigma > 0$ signals irreversible losses associated with the overall creation of correlations \cite{esp10,stra17}, while $\Delta \Sigma < 0$ reveals that work may  be gained from available correlations by reducing them \cite{fra17,man18,heg18,sal22,her23}. The condition $\Delta \Sigma < 0$ can therefore be regarded as an indicator of an entropic resource. Equation \eqref{2} provides a unified formalism enabling the investigation of both processes on the same footing.
Standard macroscopic thermodynamics completely neglects initial correlations. In that case, Eq.~\eqref{2} implies the usual second law, $\Delta S + \sum_{j} {Q_{j}}/{kT_{j}} \geq 0$ \cite{pip66}. By contrast, in the presence of initial  correlations, Eq.~\eqref{2} can lead to an inverted entropy balance, $\Delta S + \sum_{j} {Q_{j}}/{kT_{j}} \leq 0$ \cite{com}. \\

\begin{figure}[t]
    \begin{center}
    \includegraphics[scale=.56]{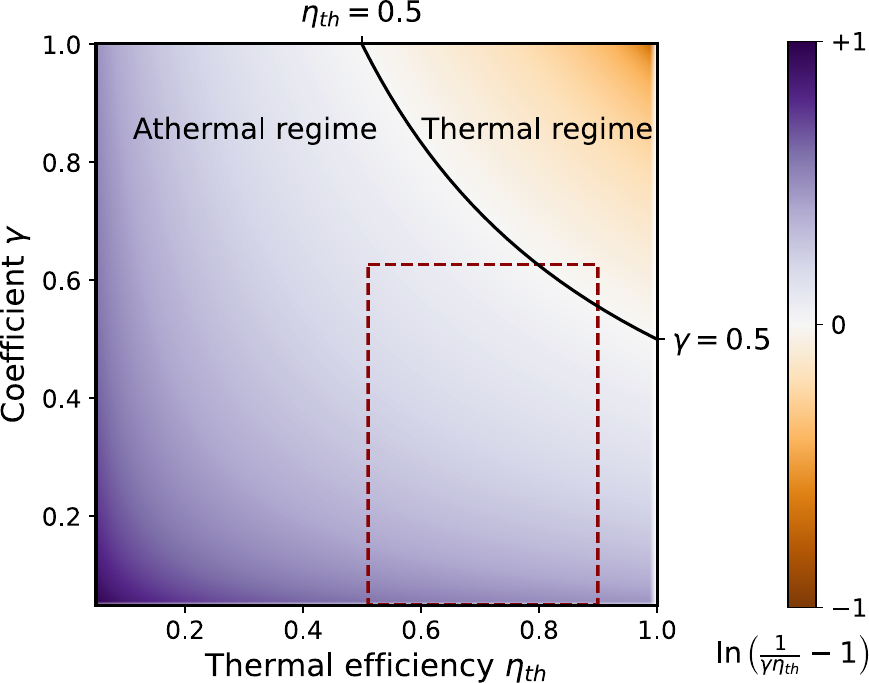}
    \caption{\textbf{Operation regimes of  correlated engines}. Depending on the value of the ratio, Eq.~\eqref{6}, of thermal and athermal contributions, correlated engines may  produce work by converting heat (thermal regime) or  entropic resources, such as correlations (athermal regime).  The plot shows the logarithm of the upper bound of Eq.~\eqref{6} (normalized to $[-1,1]$), as a function of the thermal efficiency $\eta_\text{th}$ and the coefficient $\gamma$. Engines are guaranteed to run in the thermal regime when $1 / \gamma \eta_\text{th}>2 $ (solid line). The dashed rectangle indicates the region explored by the two-oscillator-engine example of Fig.~3.}
    \label{fig:upperBound}
\end{center}
\end{figure}

 \noindent\textbf{Generalized efficiency} \\
The efficiency $\eta$,  defined  as the ratio of the  work produced by an engine and the energy needed to run it over one cycle \cite{pip66}, can be evaluated by combining the first and second laws, Eqs.~\eqref{1} and \eqref{2}. We obtain the total work for any correlated quantum engine
\begin{equation}
		W = \sum_{j} \eta_{j} Q_{j}  +  k T_\text{min}\Delta \sigma ,
	\label{eq:work}
\end{equation}
where we have introduced $k T_\text{min} \Delta \sigma =  k T_\text{min} \Delta \Sigma + \sum_{i} \Delta F_{i}$,  with $F_{i} = U_{i}  - k T_\text{min} S (\rho_{i})$  the generalized free energy \cite{esp10,stra17}. The quantity $\sigma(t+\tau)$ represents the total  entropy production during the cycle that now also accounts for thermalization processes occurring  within the parts of system when the variation $\Delta D(\rho_i || \rho_i^\text{th})\neq0$. It reduces to  $\Sigma(t+\tau)$ for an ideal  cyclic behavior of subsystem $\rho_i$ which corresponds to $\Delta F_{i}=0$. On the other hand, $\sigma(t)$ is related to all the correlations (and displacements from equilibrium) present at the beginning of a cycle. The local Carnot efficiency, $\eta_{j} = 1 - T_\text{min} / T_{j}$,  is further related to   the heat flow between the reservoir at the smallest  temperature, $T_\text{min}=\text{min}_j\{T_j\}$, and the one at temperature $T_{j}$. Contrary to conventional macroscopic heat engines, whose sole energy source is the heat $Q_j$ absorbed from bath $j$, a microscopic correlated engine might also extract work from a change of  the energy  $U$ of the system, either from a variation of the energy of subsystem $i$ or of the bath coupling energies. In small systems, these energy changes can be positive or negative along individual realizations, owing to thermal and/or quantum fluctuations. We therefore  define the corresponding contributions that originate from these two energy sources as $Q^\text{in} = \sum_{j} ( \lvert Q_{j} \rvert - Q_{j})/2$ and {$\Delta U^\text{in} = \sum_{i} ( \lvert \Delta U_{i} \rvert - \Delta U_{i})/2$, where the label \textit{in} refers to the energy that flows into the engine.} The generalized efficiency $\eta = - W /(Q^\text{in} + \Delta U^\text{in})$ then reads
\begin{equation}
	\eta = \gamma \left( \eta_\text{th} - \frac{kT_\text{min}\Delta \sigma}{Q^\text{in}}  \right),
	\label{4}
\end{equation}
where $\gamma = [ 1 + \Delta U^\text{in}/Q^\text{in}]^{-1} \leq 1$ is a parameter that assesses the fraction of  the energy used to run the  engine that comes from either the working substance or the interaction with the baths, versus the fraction that  stems from the reservoirs in the form of heat.
 
The efficiency {$\eta_\text{th} = - \sum_{j} \eta_{j} Q_{j} / Q^\text{in}$} is the thermal efficiency of a reversible  engine whose working substance is in a cyclic state with constant system-bath interactions. If all temperatures $T_{j}$ are positive, $\eta_\text{th}$ is upper bounded by the  Carnot efficiency, $\eta_\text{th} \leq \eta_\text{C} = 1 - T_\text{min} / T_\text{max}$, where $T_\text{max}$ is the largest bath temperature. We note that it is equal to $\eta_\text{C}$ if and only if all the reservoirs  from which  heat flows out  ($Q_{j} < 0$) are at temperature $T_\text{max}$, and all reservoirs from  which heat flows in ($Q_{j'} > 0$) are at temperature $T_\text{min}$. This is clearly the case for a reversible engine operating between two thermal reservoirs \cite{pip66}.

According to Eq.~\eqref{4}, the efficiency of any engine operating between many reservoirs is upper bounded  by
\begin{equation}
\label{5}
	\eta \leq  \eta_\text{C} - \frac{kT_\text{min}\Delta \sigma}{Q^\text{in}},
\end{equation}
The Carnot efficiency $\eta_\text{C}$ may hence be   surpassed  if entropic resources are present at the beginning of a cycle such that $\Delta \sigma < 0$.
 This result should not be regarded as a violation of the standard second law, but as its extension to situations where the latter does no longer apply. In standard thermodynamics, the maximum efficiency of a heat engine operating by two thermal baths is indeed given by $\eta_\text{C} - {kT_\text{min} \Sigma}/{Q^\text{in}}$ \cite{whi14}, which is always smaller than $\eta_\text{C}$ since $\Sigma \geq0$. This result follows from the  general expression \eqref{5} in the absence of initial correlations (and displacements from equilibrium). A special case of Eq.~\eqref{5} has recently been simulated experimentally using a spin-1/2 SWAP engine \cite{her23}.\\
 
 \begin{figure}[t!]
    \begin{center}
    \includegraphics[scale=.29]{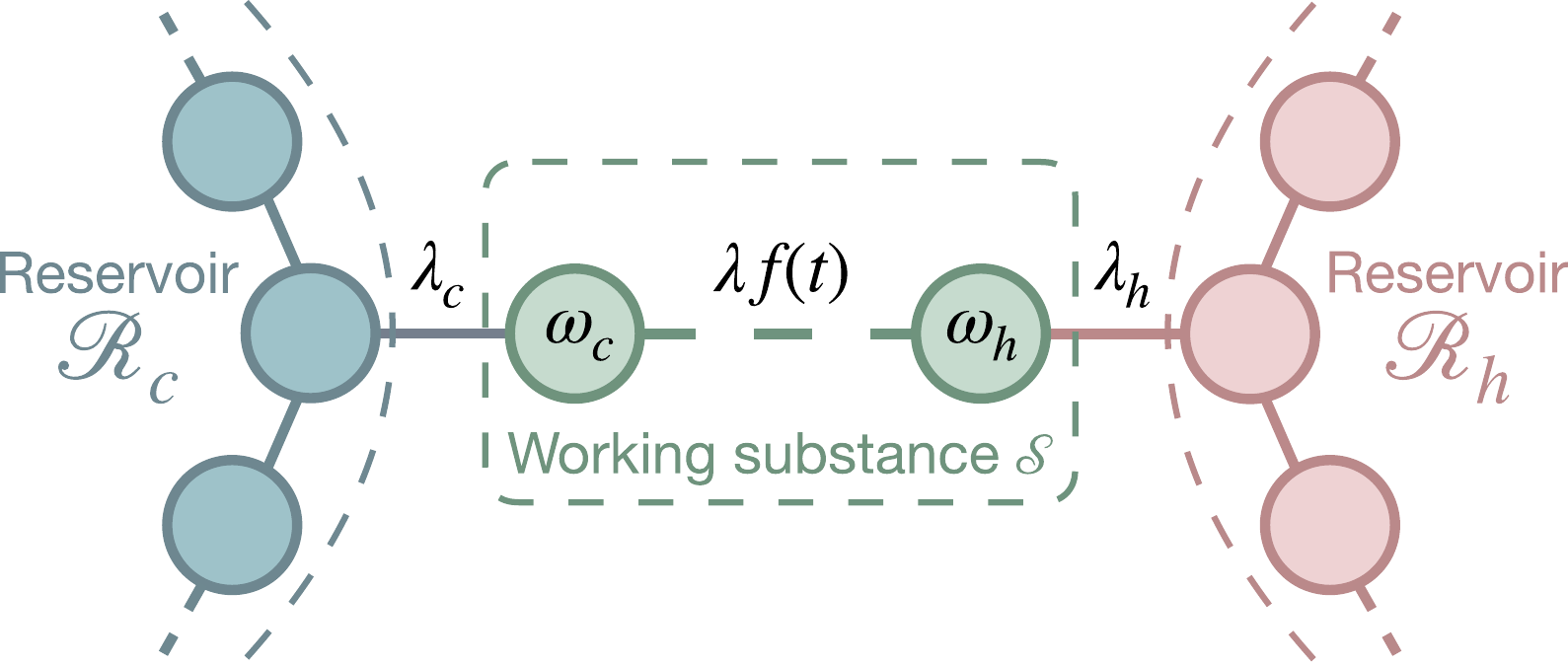}
    \caption{\textbf{Two-oscillator quantum engine}.  The working substance $\mathcal{S}$ is composed of two harmonic oscillators, with respective frequencies $\omega_\text{c}$ and $\omega_\text{h}$,  each coupled to its own reservoir, $\mathcal{R}_\text{c}$ and $\mathcal{R}_\text{h}$, with temperatures $T_\text{c}$ and  $T_\text{h}$, and coupling constants $\lambda_\text{c}$ and $\lambda_\text{h}$. Work is produced by periodically switching the oscillator interaction $\lambda f(t) $ on and off. The oscillators thermalize with their respective baths during the off phases, leading to the creation of correlations between them.}
    \label{fig:engine}
\end{center}
\end{figure}

\noindent\textbf{Characterizing entropic resources}\\
Formula \eqref{4} is exact and accounts for all possible correlations in the system-plus-baths ensemble, including displacements from equilibrium  in both system and reservoirs. It  hence allows one to  evaluate/engineer entropic resources and compute/optimize the efficiency  of any quantum engine from the knowledge of the microscopic Hamiltonians. This should be useful in the study/design of high-performance quantum engine models. Like in standard thermodynamics, where, for instance, the efficiency of an ideal Otto cycle explicitly depends on the  heat capacity of the working medium \cite{pip66}, the  numerical values of the parameters appearing in Eq.~\eqref{4}, such as $\eta$, $\eta_\text{th}$ and $\gamma$, will depend on the specific details of the considered device. However, from a macroscopic point of view, the efficiency $\eta$  follows  from the knowledge of the macroscopic energy fluxes in and out of the engine, which can, in principle, be determined experimentally \cite{her23}. Likewise, while the nonequilibrium entropy production is known to be microscopically given by the amount of correlations built between (a single) system and baths during one cycle \cite{esp10,stra17}, it is macroscopically equal to 
$\sum_{j} {Q_{j}}/{kT_{j}} = - \Sigma$ for a perfect cycle with $\sum_{i}  \Delta S (\rho_i)=0$ \cite{pip66}. In the absence of initial correlations, the entropy production can therefore be evaluated from macroscopic quantities, like temperature and heat, without having to directly determine correlation measures, such as the system-bath mutual information, which can be experimentally challenging. Similarly, still for a perfect cycle, the difference $\Delta \Sigma$ may be inferred from $\sum_{j} {Q_{j}}/{kT_{j}} = \Delta \Sigma$, in the presence of initial correlations. In the more general case of non-perfect cycles, $\sum_{i}  \Delta S (\rho_i)\neq 0$, the entropic resource $\Delta \sigma$ can be obtained from $W$, $Q_j$ and $T_j$ via Eq.~\eqref{eq:work}, again without having to assess any microscopic correlation term. In that regard, it is interesting to further note that experimental methods to detect system-reservoir correlations from measurements of the system alone are currently being developed \cite{smi11,li11,ges13,rin15}.\\

\begin{figure}[t!]
    \begin{center}
    \includegraphics[scale=.73]{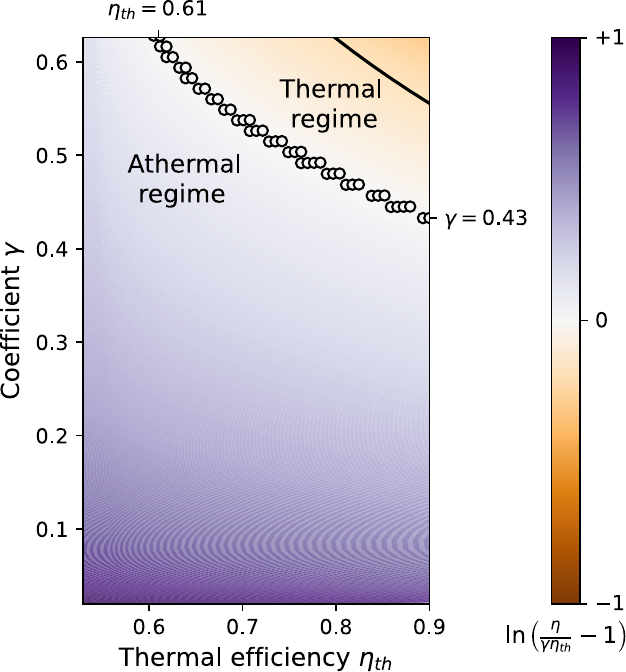}
    \caption{\textbf{Operation regimes of the two-oscillator engine}. The plot shows the (normalized) logarithm of Eq.~\eqref{6}, after  the first cycle. The black circles indicate the exact boundary between  thermal and athermal regimes given by $\eta / \gamma \eta_\text{th}=2$, whereas the solid line corresponds to the device-independent upper bound of Fig.~1. Parameters  are  $\omega_\text{c} = 1$, $\omega_\text{h} = 2$, $\lambda = 0.08$, $\lambda / 15 \leq \lambda_\text{c} = \lambda_\text{h} \leq \lambda / 2$, $T_\text{c} = 0.8$ and $1.7 \leq T_\text{h} \leq 8$.}
    \label{fig:quotientSimulation}
\end{center}
\end{figure}

\begin{figure*}[t]
    \begin{center}
    \includegraphics[scale=.42]{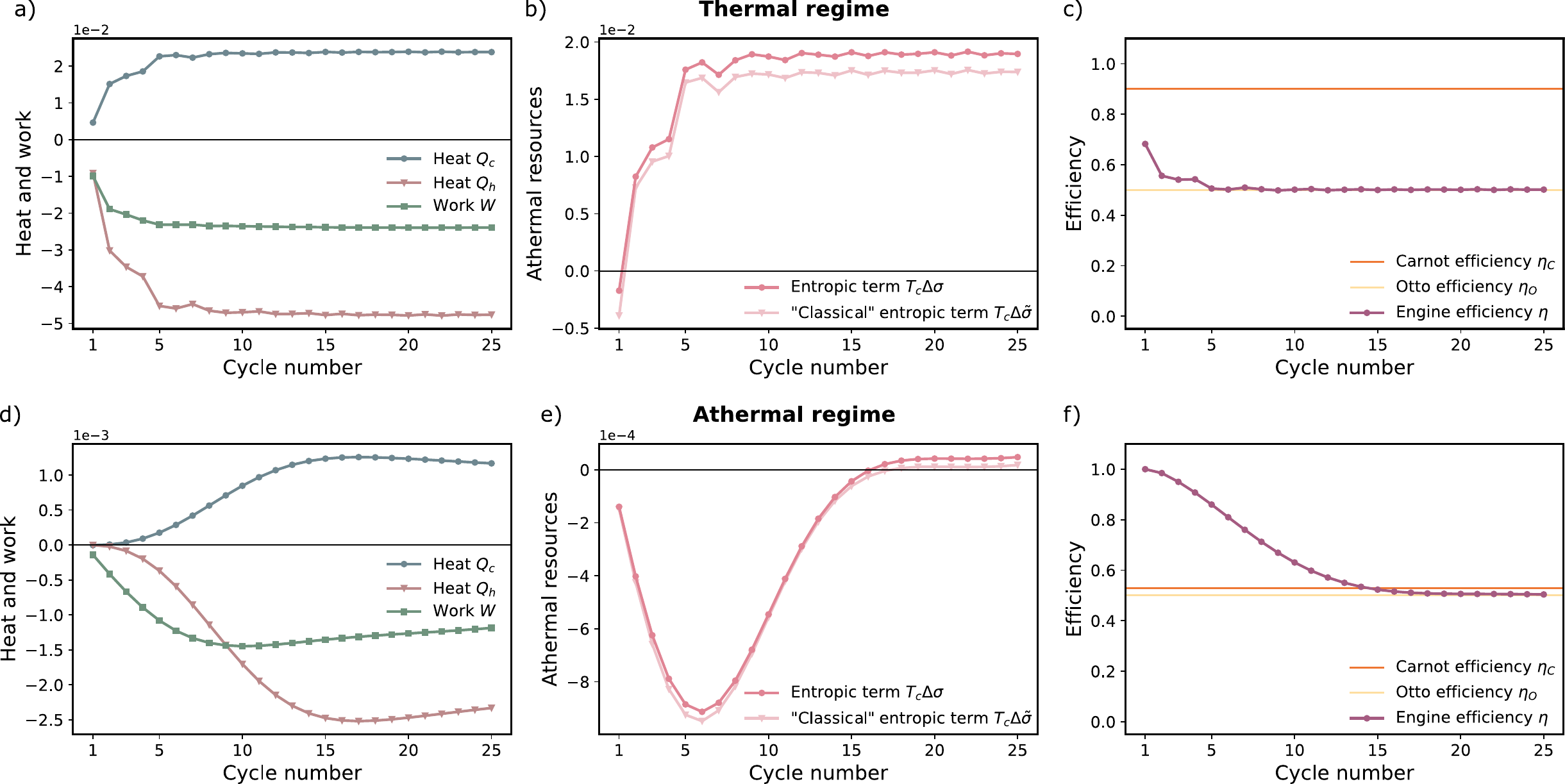}
    \caption{\textbf{Performance of the two-oscillator engine}. abc) In the thermal regime, the engine produces work from heat, while initial correlation are not exploited ($T_\text{c} \Delta \sigma >0$ after a transient).  The efficiency $\eta$ is always smaller than the Carnot efficiency $\eta_\text{C}$ and quickly converges to the Otto efficiency $\eta_\text{O}$. Parameters are $\lambda = 0.08$, $\lambda_\text{c} = \lambda_\text{h} = \lambda / 2$, $T_\text{c} = 0.8$ and $T_\text{h} = 8$ (corresponding to the upper right corner in Fig.~\ref{fig:quotientSimulation}). def) In the athermal regime, the engine predominantly produces work from  entropic resources such as system-bath correlations that are created during the thermalization with the baths: $T_\text{c} \Delta \sigma$ initially grows more negative, work is larger than the absorbed heat and the efficiency exceeds  the Carnot efficiency. As the number of cycles increases, nonequilibrium entropy produces leads to $T_\text{c} \Delta \sigma>0$, pushing  the engine to thermal operation. The 'classical' entropic term $T_\text{c} \Delta \tilde \sigma$ deviates from the fully quantum expression $T_\text{c} \Delta \sigma$. Parameters are $\lambda = 0.08$, $\lambda_\text{c} = \lambda_\text{h} = \lambda / 15$, $T_\text{c} = 0.8$ and $T_\text{h} = 1.7$ (corresponding to the lower left  corner in Fig.~\ref{fig:quotientSimulation}).}
    \label{fig:engineComparison}
\end{center}
\end{figure*}

\noindent\textbf{Athermal engine operation}\\
 Additional  insight into the physical role of entropic resources can be gained by rewriting Eq.~\eqref{4} as
\begin{equation}
	\frac{kT_\text{min} \Delta \sigma}{\sum_{j} \eta_{j} Q_{j}} = \frac{\eta}{\gamma \eta_\text{th}} - 1.
	\label{6}
\end{equation}
Equation \eqref{6} indicates that  two  regimes of engine operation should be distinguished: (1) If $\eta / \gamma \eta_\text{th}  < 2$, the engine produces work by mostly converting heat $Q_j$ from the reservoirs, that is, by exploiting  thermal resources, like common macroscopic engines. (2) {On the other hand, if $\eta / \gamma \eta_\text{th} > 2$, the engine dominantly extracts work from athermal resources, such as correlations,  displacements from equilibrium, or  interactions with the baths, quantified by $\Delta \sigma$. This new regime is possible for  correlated microscopic engines. 

In order to make model-independent statements, we next analyze the upper bound, $1 / \gamma \eta_\text{th} - 1$, of Eq.~\eqref{6}, by using the fact that the efficiency $\eta$, which depends on the details of the  engine, is always smaller than one. Figure \ref{fig:upperBound} presents the  normalized logarithm  of $1 / \gamma \eta_\text{th} - 1$ as a function of  the parameter $\gamma$ and of the thermal efficiency $\eta_\text{th}$.  The black solid line indicates the limiting value $1 / \gamma \eta_\text{th}  = 2$.  An engine is only guaranteed to run in the thermal regime   provided that $\gamma >1/2$ and $\eta_\text{th}>1/2$. The first condition is expected from the definition of the parameter $\gamma$; it simply says that the engine should be predominantly  fueled by  heat. However, surprisingly, this criterion is not enough. The nontrivial condition $\eta_\text{th} > 1/2$ reveals that, since $\eta_\text{th} \leq \eta_\text{C}$, the temperature of the hottest reservoir should  be at least twice the temperature of the colder one. Otherwise, work production could still be dominated by  athermal resources, even if $\gamma>1/2$. 

Physical intuition about the athermal regime may be developed by considering the simple example of a (single system) engine operating between two reservoirs in the weak-coupling limit. In the high-temperature (classical) domain,  thermalization of the system with a bath (either hot or cold) at the beginning of a cycle does not create any (significant) system-bath correlations. In this thermal regime, the engine is solely fueled by heat. By contrast, in the low-temperature (quantum) domain,  thermalization will  in general lead to initial system-bath correlations \cite{eis02,hil09,wil11,per11,ank14,ros18} that may be quantified by a nonzero  mutual information $I (\mathcal{S} , \mathcal{R}_j)  $. An engine will run in the athermal regime, when primarily fueled by these initial correlations, instead of heat. However, when  $\eta_\text{th} > 1/2$, that is, $T_\text{h} \geq 2 T_\text{c}$, the heat flowing between the two baths (which is proportional to the temperature difference) can become large enough that the engine is again predominantly fueled by heat, instead of  correlations.   The presence of initial correlations does hence not necessarily  imply that they will be dominantly exploited. The condition $\eta_\text{th} > 1/2$ is necessary, but not sufficient, to be in the thermal regime, as we discuss below. We additionally mention that another way to create initial system-bath correlations is to move to the strong-coupling limit.

\noindent\textbf{Two-oscillator quantum engine}\\
 As an illustration, we now consider a quantum engine whose working {substance} consists of two harmonic oscillators (with  frequencies $\omega_\text{c}$ and $
\omega_\text{h}$), each coupled to  its own reservoir ${\cal R}_\text{c}$ and ${\cal R}_\text{h}$ (with respective temperatures $T_\text{c}$ and $
T_\text{h}$ and respective coupling constants $\lambda_\text{c}$ and $\lambda_\text{h}$) (Fig.~2) \cite{k84,qua06,ajm08}. The engine is operated by periodically switching the interaction $\lambda f(t)$ between the  oscillators on and off, extracting work. {The potential $f(t)$ is taken to be a bump function with unit amplitude (Supplementary Materials). The interaction is turned on for a short time ($t_\text{on} =  41[2 \pi / (\omega_\text{h} - \omega_\text{c}) + 1]/40$) and turned off for a longer time ($t_\text{off} =  5 [2 \pi / (\omega_\text{h} - \omega_\text{c}) + 1]$) to allow the  oscillators to thermalize with their  baths. We numerically analyze the performance of the engine by modeling the finite reservoirs by an ensemble of 300 harmonic oscillators each \cite{poz18}. 
The initial state is taken to be the direct product, $\rho (0) = \rho_\text{c} \otimes \rho_\text{h}$, of the joint  thermal states $\rho_\text{c,h}$ of each oscillator with its respective reservoir. The thermalization between each oscillator and its associated bath leads to the built-up of correlations between the two \cite{eis02,hil09,wil11,per11,ank14,ros18}, implying the creation of {athermal} resources with  $\sigma(t) \neq0$ at the beginning of a cycle.

Whether the engine can exploit these {athermal} resources depends on the chosen parameter values. Figure  \ref{fig:quotientSimulation} shows the normalized logarithm of $\eta / \gamma \eta_\text{th} - 1$  when the temperature $T_\text{h}$ of the hot bath  and the  coupling constants $\lambda_\text{c}$ and $\lambda_\text{h}$ are varied in the domain where the device runs as an engine (all  other parameters are kept fixed); this, in turn, modifies the values of $\eta_\text{th}$ and $\gamma$. The athermal regime (violet) clearly dominates the thermal regime (orange), and occupies more than $87\%$ of the  parameter space; the black circles represent the numerically determined boundary between the two regimes, while the black  line is  the upper bound shown in Fig.~1.

Figure 4  further examines the performance of the  engine in the two regimes: the upper row (Figs.~4abc) displays the thermal case with $\gamma = 0.626$ and $\eta_\text{th} = 0.9$ (corresponding to the upper right corner of Fig.~3), whereas the lower row  (Figs.~4def) shows the athermal case with $\gamma = 0.02$ and $\eta_\text{th} = 0.53$ (corresponding to the lower left corner of Fig.~3). In the thermal regime, {athermal} resources are not used ({$\Delta \sigma  >0$} after a transient); the efficiency, which is always smaller than the Carnot efficiency, converges to the Otto efficiency $\eta_\text{O}=1-\omega_\text{c}/\omega_\text{h}$. By contrast, in the athermal regime, the engine initially harvests {athermal} resources ({$\Delta \sigma$} grows more negative during the first 6 cycles) and successfully converts them into work (the produced work is larger  than the absorbed heat) with an efficiency that is larger than the Carnot efficiency. This effect is mainly due to the reduction of the temperature of the hot bath, which  increases  correlations between the latter and the second oscillator after thermalization, and at the same time decreases the heat flow between the reservoirs. However, with increasing cycle number, entropy production leads to  {$\Delta \sigma>0$}, bringing the engine to  thermal operation at $\eta_\text{O}$ after 16 cycles. In this instance, the nonequilibrium entropy $\sigma(t+\tau)$ produced during one cycle via quantum friction \cite{kos02,fel03,pla14}, associated with the fact that the driving Hamiltonian does not commute with the engine Hamiltonian, becomes larger than the entropic resources $\sigma(t)$ generated at the beginning of each cycle; since overall more correlations are created than used in one cycle,  $\Delta \sigma>0$ as in the thermal regime.
The duration of the athermal regime could in principle be  significantly extended over many more cycles by employing  shortcut-to-adiabaticity techniques \cite{tor13,gue19,den13,cam14,bea16,aba17,aba18,hou25} that suppress  entropy production \cite{com1}.  It should be emphasized that the performance of the two-oscillator engine could not be described without the generalized laws (1) and (2), and the resulting efficiency (4), highlighting the importance to extend the laws of thermodynamics to correlated machines. Finally, we highlight the quantum nature of $\sigma$  by  replacing, for both system and baths,  the relative entropy  $D ( \rho || \rho^\text{th} )$ by the diagonal relative entropy $ D (   \rho_\text{d} || \rho^\text{th} )$, where $\rho_\text{d}$ is the diagonal  operator in the  energy basis \cite{com2}. The corresponding 'classical' entropic term $T_\text{c} \Delta \tilde \sigma$ deviates from the fully quantum expression $T_\text{c} \Delta  \sigma$ (Fig.~4be), indicating the presence of nonclassical correlations. \\

\noindent\textbf{DISCUSSION}\\
 The practical usefulness of the second law  is that  it provides the maximum efficiency of any thermal machine. The knowledge of such upper bound is crucial for optimization purposes \cite{pip66}. We have extended the theory of engines to the correlated quantum domain by deriving exact generalized laws of thermodynamics for cyclic processes. Since no approximations are involved, we expect them to be universally applicable. The extended   efficiency formula reveals  that there may be efficiencies above the uncorrelated Carnot limit. This is the case in the  athermal regime where engines are fueled by  entropic resources, instead of heat. As we have seen, these entropic resources can occur naturally in microscopic engines. An interesting question is to determine conditions under which correlation creation outweighs  entropy production, so that engines may remain in the athermal regime. In view of their generality,  these findings should  be relevant for the concrete design and optimization of efficient correlated quantum machines.


\clearpage
\widetext
\begin{center}
\textbf{\large Supplemental Information:  Correlated quantum machines beyond the standard second law}
\end{center}
\setcounter{equation}{0}
\setcounter{figure}{0}
\setcounter{table}{0}
\setcounter{page}{1}
\setcounter{secnumdepth}{4}
\makeatletter
\renewcommand{\theequation}{S\arabic{equation}}
\renewcommand{\thefigure}{S\arabic{figure}}
\renewcommand{\bibnumfmt}[1]{[S#1]}
\renewcommand{\citenumfont}[1]{S#1}


\renewcommand{\figurename}{Supplementary Figure}
\renewcommand{\theequation}{S\arabic{equation}}
\renewcommand{\thefigure}{S\arabic{figure}}
\renewcommand{\bibnumfmt}[1]{[S#1]}
\renewcommand{\citenumfont}[1]{S#1}

\section{Derivation of the generalized first law of thermodynamics}
\noindent In this section we derive the generalized first law of thermodynamics (Eq.~\eqref{1} of the main text):
\begin{equation}
	W = \sum_{j} Q_{j} + \Delta U,
\end{equation}
\noindent where $W = \int_{t}^{t + \tau} dt^{\prime} \langle \partial_{t^{\prime}} H_{\text{tot}} (t^{\prime}) \rangle$ is the total work extracted during a cycle of duration $\tau$, $Q_{j} = \langle H_{j} \rangle (t + \tau) - \langle H_{j} \rangle (t)$ is the heat absorbed by reservoir $\mathcal{R}_{j}$, and $\Delta U = \sum_{i} \Delta U_{i} = \sum_{i} [ \langle H_{i} + \sum_{j} H_{ij} \rangle (t + \tau) - \langle H_{i} + \sum_{j} H_{ij} \rangle (t) ]$ is  the change of the energy of the compound system, including  the interaction energy. To this end, we remind the reader that the total Hamiltonian is given by 
\begin{equation}
	H_{\text{tot}} (t) = \sum_{i} H_{i} (t) + \sum_{j} H_{j} + \sum_{i,j} H_{ij} (t),
\end{equation}
\noindent with $H_{i} (t)$ the Hamiltonian of the subsystem $\mathcal{S}_{i}$, $H_{j}$ the Hamiltonian of the reservoir $\mathcal{R}_{j}$, and $H_{ij} (t)$ the interaction Hamiltonian between $\mathcal{S}_{i}$ and $\mathcal{R}_{j}$. On the one hand, the generalized version of Ehrenfest theorem implies
\begin{equation}
	\frac{d}{dt} \langle H_{\text{tot}} \rangle = \langle \partial_{t} H_{\text{tot}} (t) \rangle.
	\label{seq:ehrenfest}
\end{equation}
\noindent On the other hand, due to the linearity of differentiation, we have
\begin{equation}
	\frac{d}{dt} \langle H_{\text{tot}} \rangle = \sum_{i} \frac{d}{dt} \langle H_{i} \rangle + \sum_{i,j} \frac{d}{dt} \langle H_{ij} \rangle + \sum_{j} \frac{d}{dt} \langle H_{j} \rangle.
	\label{seq:diff}
\end{equation}
\noindent Combining Eqs.~\eqref{seq:ehrenfest} and \eqref{seq:diff}, we further obtain

\begin{equation}
	\langle \partial_{t} H_{\text{tot}} (t) \rangle = \sum_{i} \frac{d}{dt} \langle H_{i}  + \sum_{j} H_{ij} \rangle + \sum_{j} \frac{d}{dt} \langle H_{j} \rangle.
\end{equation}
\noindent By integrating with respect to $t$, we finally arrive at the generalized first law of thermodynamics:
\begin{equation}
	\underbrace{\int_{t}^{t + \tau} dt^{\prime} \langle \partial_{t^{\prime}} H_{\text{tot}} (t^{\prime}) \rangle}_{W} = \underbrace{\sum_{i} [ \langle H_{i} + \sum_{j} H_{ij} \rangle (t + \tau) - \langle H_{i} + \sum_{j} H_{ij} \rangle (t) ]}_{\Delta U} +  \underbrace{\sum_{j} [ \langle H_{j} \rangle (t + \tau) - \langle H_{j} \rangle (t) ]}_{Q_{j}}.
	\label{seq:firstlaw}
\end{equation}
\noindent Note that
\begin{equation}
	\langle \partial_{t} H_{\text{tot}} (t) \rangle = \sum_{i} \langle \partial_{t} H_{i} (t) \rangle + \sum_{i,j} \langle \partial_{t} H_{ij} (t) \rangle
\end{equation}
\noindent and
\begin{equation}
	\frac{d}{dt} \langle H_{ij} \rangle = \frac{1}{i \hbar} \langle [ H_{ij} (t)  , H_{i} (t) + H_{j}  ] \rangle + \langle \partial_{t} H_{ij} (t) \rangle.
\end{equation}
\noindent Thus, Eq.~\eqref{seq:firstlaw} can be rewritten as
\begin{equation}
	\begin{aligned}
		\underbrace{\sum_{i} \int_{t}^{t + \tau} dt^{\prime} \langle \partial_{t^{\prime}} H_{i} (t^{\prime}) \rangle}_{W_{\mathcal{S}}}  & =  \underbrace{\sum_{j} [\langle H_{j} \rangle (t + \tau) - \langle H_{j} \rangle (t) ]}_{Q_{j}} 
		 + \underbrace{\sum_{i} [ \langle H_{i} \rangle (t + \tau) - \langle H_{i} \rangle (t) ]}_{\Delta U_{\mathcal{S}}} \\
		& + \sum_{i,j} \int_{t}^{t + \tau} dt^{\prime} \langle [ H_{ij} (t^{\prime})  , H_{i} (t^{\prime}) + H_{j}  ] \rangle.
	\end{aligned}
\end{equation}
\noindent Therefore, if $\Delta U_{\mathcal{S}} = 0$ and the interactions are energy-conserving on average, $\sum_{i,j} \int_{t}^{t + \tau} dt^{\prime} \langle [ H_{ij} (t^{\prime})  , H_{i} (t^{\prime}) + H_{j}  ] \rangle = 0$,  the standard version of the first law of thermodynamics is then recovered: $W_{\mathcal{S}} = \sum_{j} Q_{j}$.

\section{Derivation of the generalized second law of thermodynamics}

\noindent In this section we derive the generalized second law of thermodynamics (Eq.~\eqref{2} of the main text):
\begin{equation}
	\sum_{i} \Delta S ( \rho_{i}) + \sum_{j} \frac{Q_{j}}{k T_{j}} = \Delta \Sigma,
\end{equation}
\noindent where $k$ is the Boltzmann constant, $\rho_{i} = \text{tr}_{\bar{\mathcal{S}}_{i}} (\rho_{\mathcal{SR}})$ is the reduced state of the subsystem $\mathcal{S}_{i}$ ($\bar{\mathcal{S}}_{i}$ is the complement of $\mathcal{S}_{i}$), $\Delta S (\rho_{i}) = S(\rho_{i} (t + \tau)) - S(\rho_{i} (t))$ with $S (\rho) = - \text{tr} [\rho \, \text{ln} (\rho)]$ the von Neumann entropy, and $\Delta \Sigma = \Sigma (t + \tau) - \Sigma (t)$ with $\Sigma = I (\mathcal{S} , \mathcal{R}) + C (\mathcal{S}) + C (\mathcal{R}) + \sum_{j} D ( \rho_{j} || \rho_{j}^{\text{th}} )$.  Additionally, $I (\mathcal{S} , \mathcal{R}) = S(\rho_{\mathcal{S}}) + S(\rho_{\mathcal{R}}) - S (\rho_{\mathcal{SR}})$ is the mutual information between the system $\mathcal{S}$ and the collection of all reservoirs $\mathcal{R}$, and $C (\mathcal{S}) = \sum_{i} S(\rho_{i}) - S(\rho_{\mathcal{S}})$ and $C (\mathcal{R}) = \sum_{j} S(\rho_{j}) - S(\rho_{\mathcal{R}})$ are the total correlations between all the subsystems and all the reservoirs, respectively. Furthermore, $D ( \rho_{j} || \rho_{j}^{\text{th}} )$ is the relative entropy between the reduced state of $\mathcal{R}_{j}$, $\rho_{j}$, and the reference thermal state $\rho_{j}^{\text{th}} = \text{exp}( - H_{j} / k T_{j}) / Z_{j}$. We begin by showing the validity of the following identity:
\begin{equation}
	\begin{aligned}
		\sum_{i} \Delta S (\rho_{i}) + \sum_{j} \Delta S (\rho_{j}) & = \Delta I (\mathcal{S} , \mathcal{R}) + \Delta C (\mathcal{S}) + \Delta C (\mathcal{R}) \\
		& = \underbrace{\left[ \Delta S (\rho_{\mathcal{S}}) + \Delta S (\rho_{\mathcal{R}}) - \Delta S (\rho_{\mathcal{S R}}) \right]}_{\Delta I (\mathcal{S} , \mathcal{R})} 
		 + \underbrace{\left[ \sum_{i} \Delta S (\rho_{i}) - \Delta S (\rho_{\mathcal{S}}) \right]}_{\Delta C (\mathcal{S})} + \underbrace{\left[ \sum_{j} \Delta S (\rho_{j}) - \Delta S (\rho_{\mathcal{R}}) \right]}_{\Delta C (\mathcal{R})} \\
		& =  \sum_{i} \Delta S (\rho_{i}) + \sum_{j} \Delta S (\rho_{j}) 
		 + \underbrace{[\Delta S (\rho_{\mathcal{S}}) - \Delta S (\rho_{\mathcal{S}})]}_{= 0} + \underbrace{[\Delta S (\rho_{\mathcal{R}}) - \Delta S (\rho_{\mathcal{R}})]}_{=0} - \underbrace{\Delta S (\rho_{\mathcal{S R}})}_{=0} \\
		& = \sum_{i} \Delta S (\rho_{i}) + \sum_{j} \Delta S (\rho_{j}).
	\end{aligned}
	\label{seq:identity}
\end{equation}
\noindent We next write the relative entropy $D ( \rho_{j} || \rho_{j}^{\text{th}} )$ explicitly as
\begin{equation}
	\begin{aligned}
		D ( \rho_{j} (t) || \rho_{j}^{\text{th}} ) & = \text{tr} [\rho_{j} \, \text{ln} (\rho_{j})] - \text{tr} [\rho_{j} \, \text{ln} (\rho_{j}^{\text{th}})] 
		 = - S (\rho_{j}) + \frac{1}{k T_{j}} \text{tr} [\rho_{j} (t) H_{j}] + \text{ln} (Z_{j}),
	\end{aligned}
\end{equation}
\noindent which implies
\begin{equation}
	\Delta S (\rho_{j}) = \frac{Q_{j}}{k T_{j}} - \Delta D ( \rho_{j} || \rho_{j}^{\text{th}} ).
	\label{seq:freeEnergy}
\end{equation}
\noindent Finally, we replace Eq.~\eqref{seq:freeEnergy} in the identity in Eq.~\eqref{seq:identity} to obtain the generalized second law of thermodynamics:
\begin{equation}
	\sum_{i} \Delta S ( \rho_{i}) + \sum_{j} \frac{Q_{j}}{k T_{j}} =  \underbrace{\Delta I (\mathcal{S} , \mathcal{R}) + \Delta C (\mathcal{S}) + \Delta C (\mathcal{R}) + \sum_{j} \Delta D ( \rho_{j} || \rho_{j}^{\text{th}} )}_{\Delta \Sigma}.
\end{equation}

\section{Derivation of the generalized efficiency}
\noindent  Let us next derive a generalized formula for the efficiency of any quantum engine (Eq.~\eqref{4} of the main text):
\begin{equation}
	\eta = \gamma \left( \eta_{\text{th}} - \frac{k T_{\text{min}} \Delta \sigma}{Q^{\text{in}}} \right),
\end{equation}
\noindent where $\gamma = [1 + \Delta U^{\text{in}}/Q^{\text{in}}]^{-1}$ and $\eta_{\text{th}} = - \sum_{j} \eta_{j} Q_{j} / Q^{\text{in}}$, with $Q^{\text{in}} = \sum_{j} ( \lvert Q_{j} \rvert - Q_{j})/2$ and $\Delta U^{\text{in}} = \sum_{i} ( \lvert \Delta U_{i} \rvert - \Delta U_{i})/2$. To this end, we first rewrite the first law of thermodynamics as
\begin{equation}
	\begin{aligned}
		W & =  \sum_{j} Q_{j} + \sum_{i} k T_{\text{min}} \Delta S(\rho_{i}) + \sum_{i} [ \Delta U_{i} - k T_{\text{min}} \Delta S(\rho_{i}) ]  \\
		    & = \sum_{j} Q_{j} + \sum_{i} k T_{\text{min}} \Delta S(\rho_{i}) + \sum_{i} \Delta F_{i}
	\end{aligned}
\end{equation}
with $F_{i} = U_{i}  - k T_\text{min} S (\rho_{i})$  the generalized free energy. On the other hand, from the second law of thermodynamics we know
\begin{equation}
	\sum_{i} \Delta S ( \rho_{i}) = - \sum_{j} \frac{Q_{j}}{k T_{j}} + \Delta \Sigma.
\end{equation}
We now combine the two previous equations to obtain Eq.~\eqref{eq:work} of the main text:
\begin{equation}
		W = \sum_{j} \underbrace{\left( 1 - \frac{T_{\text{min}}}{T_{j}} \right)}_{\eta_{j}} Q_{j} + \underbrace{k T_{\text{min}} \Delta \Sigma + \sum_{i} \Delta F_{i}}_{k T_{\text{min}} \Delta \sigma}
	\label{seq:firstAndSecond}
\end{equation}
\noindent Finally, we compute the efficiency $\eta$ with the help of Eq.~\eqref{seq:firstAndSecond} and find
\begin{equation}
	\begin{aligned}
		\eta & = - \frac{W}{Q^{\text{in}} + \Delta U^{\text{in}}} 
		 = - \underbrace{\frac{1}{1 + \frac{ \Delta U^{\text{in}}}{Q^{\text{in}}}}}_{\gamma} \frac{W}{Q^{\text{in}}}  = \gamma \left( \underbrace{- \sum_{j} \eta_{j} \frac{Q_{j}}{Q^{\text{in}}}}_{\eta_{\text{th}}}  - \frac{k T_{\text{min}} \Delta \sigma}{Q^{\text{in}}} \right).
	\end{aligned}
\end{equation}

\section{Details of the numerical simulations}
\noindent The Hamiltonian of the two-oscillator engine simulated  in the main text is
\begin{equation}
	\begin{aligned}
		H_{tot} (t)  & = \frac{1}{2} \left(  p_{c}^{2} +  \omega_{c}^{2} x_{c}^{2} \right)  +  \frac{1}{2} \left(  p_{h}^{2} +  \omega_{h}^{2} x_{h}^{2} \right) + \lambda f(t) x_{c} x_{h} \\
		 &+  \sum_{\alpha=1,\dots,300} \frac{1}{2} \left( \pi_{c, \alpha}^{2} +  \omega_{c}^{2} q_{c, \alpha}^{2} + 2\lambda_{c} q_{c, \alpha} q_{c, \alpha + 1} \right) + \lambda_{c} x_{c} q_{c, 1} \\
		 &+  \sum_{\alpha=1,\dots,300} \frac{1}{2} \left( \pi_{h, \alpha}^{2} +  \omega_{h}^{2} q_{h, \alpha}^{2} + 2\lambda_{h} q_{h, \alpha} q_{h, \alpha + 1} \right) + \lambda_{h} x_{h} q_{h, 1},
	\end{aligned}
\end{equation}
\noindent where $p_{c,h}$ and $x_{c,h}$ are the momentum and position operators of the harmonic oscillators composing the working substance $\mathcal{S}$, and $\pi_{c(h), \alpha}$ and $q_{c(h), \alpha}$ the ones corresponding to the oscillators of the cold (hot) reservoir $\mathcal{R}_{c}$ ($\mathcal{R}_{h}$). The frequencies and the couplings were assigned the values $\omega_{c} = 1$, $\omega_{h} = 2$, $\lambda = 0.08$, $\lambda / 15 \leq \lambda_{c} = \lambda_{h} \leq \lambda /2$. Note that we used periodic boundary conditions for the reservoirs: $q_{c, 301} = q_{c, 1}$ and the analogous for the other operators. The bump function $f(t)$ modulating the interaction between the two oscillators of $\mathcal{S}$ is given by
\begin{equation}
	f(t) =
	\begin{dcases}
	\frac{1}{2} \left\{1 - \text{tanh} \left[ \text{cot} \left( \pi t / \delta \right) \right] \right\} \quad & 0 \leq t \leq \delta \\
	1 \quad & \delta < t < t_{on} - \delta \\
	\frac{1}{2} \left\{1 + \text{tanh} \left\{ \text{cot} \left[ \pi (t - t_{on}) / \delta \right] \right\} \right\} \quad & t_{on} - \delta \leq t \leq t_{on} \\
	0 \qquad & t_{on} < t \leq t_{on} + t_{off},
	\end{dcases}
\end{equation}
and extended periodically to $t > t_{on} + t_{off}$ with period $\tau = t_{on} + t_{off}$. The on and off times used were $t_{on} = 41[2 \pi / (\omega_h - \omega_c) + 1]/40$ and $t_{off} = 5 [2 \pi / (\omega_h - \omega_c) + 1]$. Additionally, $\delta$ is the time it takes to fully switch on or off the interaction between the two osillators and was given by $\delta = 0.45 \, t_{on}$. The global initial state of the engine was $\rho(t = 0) = \rho_{c} \otimes \rho_{h}$, where
\begin{equation}
	\rho_{c} = \text{exp} \left\{ - \left[  \frac{1}{2} \left(  p_{c}^{2} +  \omega_{c}^{2} x_{c}^{2} \right)  +  \sum_{\alpha=1,\dots,300} \frac{1}{2} \left( \pi_{c, \alpha}^{2} +  \omega_{c}^{2} q_{c, \alpha}^{2} + 2\lambda_{c} q_{c, \alpha} q_{c, \alpha + 1} \right) + \lambda_{c} x_{c} q_{c, 1} \right] / T_{c} \right\} / Z_{c}
\end{equation}
\noindent and
\begin{equation}
	\rho_{h} = \text{exp} \left\{ - \left[  \frac{1}{2} \left(  p_{h}^{2} +  \omega_{h}^{2} x_{h}^{2} \right)  +  \sum_{\alpha=1,\dots,300} \frac{1}{2} \left( \pi_{h, \alpha}^{2} +  \omega_{h}^{2} q_{h, \alpha}^{2} + 2\lambda_{h} q_{h, \alpha} q_{h, \alpha + 1} \right) + \lambda_{h} x_{h} q_{h, 1} \right] / T_{h} \right\} / Z_{h},
\end{equation}
\noindent with $T_{c} = 0.8$, $1.7 \leq T_{h} \leq 8$, and $Z_{c,h}$ the corresponding normalizations.

\noindent The simulations were done using Python with a modified version of the code used in Ref.~\cite{poz18}.

\section{Comparison with a previous approach}
\noindent It is instructive to compare our exact findings with those obtained using  different approaches. We shall concretely compare our generalized second law with the generalized Clausius inequality for correlated systems obtained using an information-theoretic approach  in  
Ref.~\cite{ber17} (see Eq.~(8) in that article).  This result is based on a number of assumptions,  in particular, an entropy preserving operation between two infinite-sized thermal baths $A$ and $B$ which leaves their respective Hamiltonians unchanged. Assuming further that  such operation is energy non-increasing and that the initial state of the baths is an uncorrelated product of thermal states,  the heat flow between the baths is found to obey the generalized Clausius inequality 
\begin{equation}
\label{seq:ber}
- \tilde{Q}_{A} (T_{B} - T_{A}) \geq k T_{A} T_{B} \Delta I (\mathcal{R}_{A} , \mathcal{R}_{b}),
\end{equation}
\noindent where $T_{A,B}$ are the initial bath temperatures, $I (\mathcal{R}_{A} , \mathcal{R}_{b})$ is the mutual information between them, and   $\tilde{Q}_{A} = - k T_{A} \Delta S (\rho_{A})$ is the heat (with $S$ the von Neumann entropy). \\

\noindent Equation \eqref{seq:ber} can be recovered, and extended, starting from our exact first and second laws, Eqs.~(1) and (2) of the main text.
We model direct contact between the two reservoirs by considering a working substance whose state is cyclic, whose Hamiltonian is constant in time, and that does not get correlated with the two baths. In that way, the working substance does not disturb the states of the baths, and only acts as a connection between them. Setting moreover $\Delta U_{\text{int}} = 0$, we find from our generalized first and second laws
\begin{equation}
	Q_{A} (T_{B} - T_{A}) = k T_{A} T_{B} \left[ \Delta I (\mathcal{R}_{A} , \mathcal{R}_{b}) +  D ( \rho_{A} || \rho_{A}^{\text{th}} ) + D ( \rho_{B} || \rho_{B}^{\text{th}} )  \right].
\end{equation}
\noindent We first observe that, since the relative entropies are positive, we have $Q_{A} (T_{B} - T_{A}) \geq k T_{A} T_{B} \Delta I (\mathcal{R}_{A} , \mathcal{R}_{A})$. We hence recover an inequality for the heat that is similar to that of Ref.~\cite{ber17}. However, the two definitions of heat differ: ours is based on energy changes, while that of Ref.~\cite{ber17} is based on entropy changes. Both definitions can be related by noting that, since the initial state of the reservoirs is thermal, $D ( \rho_{A} || \rho_{A}^{\text{th}} ) = \Delta D ( \rho_{A} || \rho_{A}^{\text{th}} ) = Q_{A} / k T_{A} - \Delta S (\rho_{A}) = (Q_{A}+ \tilde{Q}_{A})/kT_{A}$. In other words,   $ Q_{A}=- \tilde{Q}_{A} + kT_{A} \Delta D ( \rho_{A} || \rho_{A}^{\text{th}} )$, showing that both expressions agree at equilibrium (except for the sign). Using the above equality and adding $T_{A} \tilde{Q}_{A}$ to both sides, we obtain the exact equality
\begin{equation}
	- \tilde{Q}_{A} (T_{B} - T_{A}) = k T_{A} T_{B} \Delta I (\mathcal{R}_{A} , \mathcal{R}_{b}) + k T_{A} \left[ T_{A} \Delta D ( \rho_{A} || \rho_{A}^{\text{th}} ) + T_{B} \Delta D ( \rho_{B} || \rho_{B}^{\text{th}} )  \right],
\end{equation}
\noindent which generalizes the inequality \eqref{seq:ber} of Ref.~\cite{ber17}.


\begin{thebibliography}{99}

\bibitem{pip66} M. J. Moran and H. N. Shapiro, \textit{Fundamentals of Engineering Thermodynamics}, (Wiley, Chichester, 2006).
\bibitem{lan76} L. D. Landau and E. M. Lifschitz, \textit{Statistical Physics}, (Pergamon Press, Oxford, 1976). 

\bibitem{tal20} P. Talkner and P. H\"anggi, Colloquium: Statistical mechanics and thermodynamics at strong coupling: Quantum and classical, Rev. Mod. Phys. \textbf{92}, 041002 (2020).
\bibitem{lan21} G. T. Landi and M. Paternostro, Irreversible entropy production: From classical to quantum, Rev. Mod. Phys. \textbf{93}, 035008 
(2021). 

\bibitem{eis02} J. Eisert and M. B. Plenio, Quantum and Classical Correlations in Quantum Brownian Motion, Phys. Rev. Lett. \textbf{89}, 137902 (2002).
\bibitem{hil09} S. Hilt and E. Lutz, System-bath entanglement in quantum thermodynamics, Phys. Rev. A \textbf{79}, 010101(R) (2009).
\bibitem{wil11} N. S. Williams, K. Le Hur and A. N. Jordan, Effective thermodynamics of strongly coupled qubits, J. Phys. A: Math. Theor. \textbf{44}, 385003 (2011).
\bibitem{per11} A. Pernice and W. T. Strunz, Decoherence and the nature of system-environment correlations, Phys. Rev. A \textbf{84}, 062121 (2011).
\bibitem{ank14} J. Ankerhold and J. P. Pekola, Heat due to system-reservoir correlations in thermal equilibrium, Phys. Rev. B \textbf{90}, 075421 (2014).
\bibitem{ros18} K. Roszak, Criteria for system-environment entanglement generation for systems of any size in pure-dephasing evolutions, Phys. Rev. A \textbf{98}, 052344 (2018).

\bibitem{all00} A. E. Allahverdyan, T. M. Nieuwenhuizen, Extraction of work from a single thermal bath in the quantum regime, Phys. Rev. Lett.
 \textbf{85}, 1799 (2000).
 \bibitem{car16} M. Carrega, P. Solinas, M. Sassetti, and U. Weiss, Energy Exchange in Driven Open Quantum Systems at Strong Coupling, Phys. Rev. Lett. \textbf{116}, 240403 (2016).
\bibitem{ber17} M. N. Bera, A. Riera, M. Lewenstein, and A. Winter,  Generalized laws of thermodynamics in the presence of correlations, Nature Comm. \textbf{8},  2180 (2017).

\bibitem{per18} M. Perarnau-Llobet, H. Wilming, A. Riera, R. Gallego, and J. Eisert, Strong Coupling Corrections in Quantum Thermodynamics, Phys. Rev. Lett. \textbf{120}, 120602 (2018).
\bibitem{str19} P. Strasberg, Repeated Interactions and Quantum Stochastic Thermodynamics at Strong Coupling, Phys. Rev. Lett. \textbf{123}, 180604 (2019).
\bibitem{mic19} K. Micadei, J. P. S.  Peterson, A. M. Souza,  R. S. Sarthour, I.  S. Oliveira, G. T. Landi, T. B. Batalhao, R. M. Serra, and E. Lutz, 
 Reversing the direction of heat flow using quantum correlations, Nature Comm. \textbf{10},   2456 (2019).
 \bibitem{lip24} P. Lipka-Bartosik, G. F. Diotallevi, and P. Bakhshinezhad, Fundamental Limits on Anomalous Energy Flows in Correlated Quantum Systems, Phys. Rev. Lett. \textbf{132}, 140402 (2024).

 \bibitem{sap19} F. Sapienza, F. Cerisola, and A. J. Roncaglia, Correlations as a resource in quantum thermodynamics, Nature Comm. \textbf{10},  2492 (2019).
\bibitem{riv20} A. Rivas, Strong Coupling Thermodynamics of Open Quantum Systems, Phys. Rev. Lett. \textbf{124}, 160601 (2020).
\bibitem{cre21} J. D. Cresser and J. Anders, Weak and Ultrastrong Coupling Limits of the Quantum Mean Force Gibbs State, Phys. Rev. Lett. \textbf{127}, 250601 (2021).
\bibitem{liu21} J. Liu, K. A. Jung, and D. Segal, Periodically Driven Quantum Thermal Machines from Warming up to Limit Cycle, Phys. Rev. Lett. \textbf{127}, 200602 (2021).



\bibitem{kos13} R. Kosloff, Quantum Thermodynamics: A Dynamical Viewpoint, Entropy \textbf{15}, 2100 (2013).
\bibitem{kos14} R. Kosloff and A. Levy, Quantum Heat Engines and Refrigerators: Continuous Devices, Annu. Rev. Phys. Chem. \textbf{65}, 365 (2014).
\bibitem{ben17} G. Benenti, G. Casati, K. Saito, and R. S. Whitney, Fundamental aspects of steady-state conversion of heat to work at the nanoscale, Phys. Rep. \textbf{694}, 1 (2017).
\bibitem{mye22} N. M. Myers, O. Abah, and S. Deffner, Quantum thermodynamic devices: from theoretical proposals to experimental reality, AVS Quantum Sci. \textbf{4}, 027101 (2022).
\bibitem{arr23} L. Arrachea, Energy dynamics, heat production and heat-work conversion with qubits: toward the development of quantum machines, Rep. Prog. Phys. \textbf{86}, 036501 (2023).
\bibitem{can24}  L. M. Cangemi, C. Bhadra, A. Levy, Quantum engines and refrigerators, Phys. Rep. \textbf{1087}, 1 (2024).

\bibitem{par08} Partovi, M. H., Entanglement versus stosszahlansatz:
disappearance of the thermodynamic arrow in a high-correlation environment.
{Phys. Rev. E} \textbf{77}, 021110 (2008).

\bibitem{jen10} Jennings, D. \& Rudolph, T. Entanglement and the thermodynamic
arrow of time. {Phys. Rev. E} \textbf{81}, 061130 (2010).






\bibitem{scu03} M. O. Scully, M. S. Zubairy, G. Agarwal, and H. Walther, Extracting Work from a Single Heat Bath via Vanishing Quantum Coherence, Science \textbf{299}, 862 (2003).
\bibitem{scu11} M. O. Scully, K. R. Chapin, K. E. Dorfman, M. B.
Kim, and Svidzinsky,  Quantum heat engine power can be increased by noise-induced coherence, Proc. Natl. Acad. Sci. U.S.A. \textbf{108}, 15097 (2011).

\bibitem{dil09} R. Dillenschneider and E. Lutz, Energetics of quantum correlations, EPL \textbf{88},  50003 (2009).
\bibitem{per15} M. Perarnau-Llobet, K. V. Hovhannisyan, M. Huber, P. Skrzypczyk, N. Brunner, and A. Ac\'in, Extractable Work from Correlations, Phys. Rev. X \textbf{5}, 041011 (2015).

\bibitem{ros14} J.  Ro\ss nagel, O.  Abah, F.  Schmidt-Kaler, K.  Singer, and E. Lutz,  Nanoscale heat engine
beyond the Carnot limit. Phys. Rev. Lett. \textbf{112}, 030602 (2014).
\bibitem{nie18} W.  Niedenzu, V.  Mukherjee, A. Ghosh, A. G. Kofman, and G.  Kurizki,  Quantum engine
efficiency bound beyond the second law of thermodynamics, Nat. Commun. \textbf{9}, 165 (2018).

\bibitem{k84}
R. Kosloff, {A quantum mechanical open system as a model of a heat engine,} J. Chem. Phys. \textbf{80}, 1625 (1984).

\bibitem{qua06} H. T. Quan, Y. D. Wang, Y. X. Liu, C. P. Sun, and F. Nori, Maxwell's Demon Assisted Thermodynamic Cycle in Superconducting Quantum Circuits, Phys. Rev. Lett. \textbf{97}, 180402 (2006).

\bibitem{ajm08}
A. Allahverdyan, R. Johal, and G. Mahler, {Work extremum principle: Structure and function of quantum heat engines}, Phys. Rev. E \textbf{77}, 041118 (2008).
\bibitem{lbplb2023}
P. Lipka-Bartosik, M. Perarnau-Llobet, and N. Brunner, {Operational Definition of the Temperature of a Quantum State}, Phys. Rev. Lett. \textbf{130}, 040401 (2023).

\bibitem{cic22} F. Ciccarello, S. Lorenzo, V. Giovannetti, and G. M. Palma, Quantum collision models: Open system dynamics from repeated interactions, Phys. Rep. \textbf{954}, 1 (2022).

\bibitem{esp10} M. Esposito, K. Lindenberg, and C. Van den Broeck, Entropy production as correlation between system and reservoir, New J. Phys. \textbf{12}, 013013 (2010).
\bibitem{stra17} P. Strasberg, G. Schaller,  T. Brandes, and M. Esposito, Quantum and Information Thermodynamics: A Unifying Framework Based
on Repeated Interactions, Phys. Rev. X  \textbf{7}, 021003 (2017).

\bibitem{pus78} W. Pusz and S. L. Woronowicz, Passive states and KMS states for general quantum systems, Commun.Math. Phys. \textbf{58}, 273 (1978).

\bibitem{bru16} A. Bruch, M. Thomas, S. Viola Kusminskiy, F. von Oppen,
and A. Nitzan, Quantum thermodynamics of the driven resonant level model, Phys. Rev. B \textbf{93}, 115318 (2016).
\bibitem{new17} D. Newman, F. Mintert, and A. Nazir, Performance of a quantum heat engine at strong reservoir coupling, Phys. Rev. E \textbf{95}, 032139 (2017).
\bibitem{wie20} M. Wiedmann, J. T. Stockburger, and J. Ankerhold, Non-Markovian dynamics of a quantum heat engine: Out-of-
equilibrium operation and thermal coupling control, New J. Phys. \textbf{22}, 033007 (2020).
\bibitem{can21} L. M. Cangemi, M. Carrega, A. De Candia, V. Cataudella, G. De Filippis, M. Sassetti, and G. Benenti, Optimal energy conversion through antiadiabatic driving breaking time-reversal symmetry, Phys. Rev. Research \textbf{3}, 013237 (2021).

\bibitem{aul09} G. Auletta, M. Fortunato and G. Parisi, \textit{Quantum Mechanics}, (Cambridge University Press, Cambridge, 2009)

\bibitem{cov91} T. M. Cover and J. A. Thomas, \textit{Elements of Information Theory}, (Wiley, New York, 1991).

\bibitem{w1960}
S. Wantanabe, Information Theoretical Analysis of Multivariate Correlation, IBM J. Res. Dev. \textbf{4}, 66 (1960).

\bibitem{mod12} K. Modi, A. Brodutch, H. Cable, T. Paterek, and V. Vedral, The classical-quantum boundary for correlations: Discord and related measures, Rev. Mod. Phys. \textbf{84}, 1655 (2012).

\bibitem{ond81} M. J. Ondrechen, B. Andresen, M. Mozurkewich, and R. S. Berry, Maximum work from a finite reservoir by sequential Carnot cycles, Am. J.  Phys. \textbf{49}, 681 (1981).
\bibitem{wan16} Y. Wang, Optimizing work output for finite-sized heat reservoirs: Beyond linear response, Phys. Rev. E \textbf{93}, 012120 (2016).
\bibitem{taj17} H. Tajima and M. Hayashi, Finite-size effect on optimal efficiency of heat engines, Phys. Rev. E \textbf{96}, 012128 (2017).
\bibitem{poz18} A. Pozas-Kerstjens, E. G. Brown and K. V. Hovhannisyan, A quantum Otto engine with finite heat baths: energy, correlations, and degradation, New J. Phys. \textbf{20}, 043034 (2018).
\bibitem{moh19} M. H. Mohammady and A. Romito,
 Efficiency of a cyclic quantum heat engine with finite-size baths,
Phys. Rev. E \textbf{100}, 012122 (2019).
\bibitem{ma20} Y. H. Ma, Effect of Finite-Size Heat Source's Heat Capacity on the Efficiency of Heat Engine, Entropy \textbf{22}, 1002 (2020).


\bibitem{fra17} G. Francica, J. Goold, F.  Plastina, and M. Paternostro, Daemonic ergotropy: enhanced work extraction from quantum correlations, npj Quantum Inf. \textbf{3}, 12 (2017).
\bibitem{man18} G. Manzano, F. Plastina, and R. Zambrini, Optimal Work Extraction and Thermodynamics of Quantum Measurements and Correlations, Phys. Rev. Lett. \textbf{121}, 120602 (2018).
\bibitem{heg18} A. Hewgill, A. Ferraro, and G. De Chiara, Quantum correlations and thermodynamic performances of two-qubit engines with local and common baths, Phys. Rev. A \textbf{98}, 042102 (2018).
\bibitem{sal22} R. Salvia and V. Giovannetti, Extracting work from correlated many-body quantum systems, Phys. Rev. A \textbf{105}, 012414 (2022).
\bibitem{her23}
M. Herrera, J. Reina, I. D'Amico, and R. Serra, Correlation-boosted quantum engine: A proof-of-principle demonstration, Phys. Rev. Research \textbf{5}, 043104 (2023).

\bibitem{com} Note that an inverted entropy balance also occurs  for initially isolated equilibrium states that dissipatively evolve into a nonthermal state \cite{par89,per02,gav16,may23}.
\bibitem{par89} Partovi, M. H. Quantum thermodynamics, {\it Phys. Lett. A} \textbf{137}, 440 (1989).
\bibitem{per02}  Peres, A. \textit{Quantum Theory: Concepts and Methods}, (Kluwer Academic Publishers, New York, 2002), Chap.~9.
\bibitem{gav16}  Gaveau, B.,  Granger, L., Moreau, M. \& Schulman L. S. , Relative Entropy, Interaction Energy and the Nature of Dissipation, {\it Entropy} \textbf{16}, 3173 (2016).
\bibitem{may23} D. Mayer, E. Lutz, and A. Widera, Generalized Clausius inequalities in a nonequilibrium cold-atom system, Commun. Phys. \textbf{6}, 61 (2023).


\bibitem{whi14} R. S. Whitney, Most Efficient Quantum Thermoelectric at Finite Power Output, Phys. Rev. Lett. \textbf{112}, 130601 (2014).

\bibitem{smi11} A. Smirne, D. Brivio, S. Cialdi, B. Vacchini, and M. G. A. Paris, Experimental investigation of initial system-environment correlations via trace-distance evolution, Phys. Rev. A \textbf{84}, 032112 (2011).
\bibitem{li11} C.-F. Li, J.-S. Tang, Y.-L. Li, and G.-C. Guo, Experimentally witnessing the initial correlation between an open quantum system and its environment, Phys. Rev. A \textbf{83}, 064102 (2011).
\bibitem{ges13} M. Gessner, M. Ramm, T. Pruttivarasin, A. Buchleitner, H.-P. Breuer, and H. H\"affner, Local detection of quantum correlations with a single trapped ion, Nat. Phys. \textbf{10}, 105 (2013).
\bibitem{rin15} M. Ringbauer, C. J. Wood, K. Modi, A. Gilchrist, A. G. White, and A. Fedrizzi,
Characterizing Quantum Dynamics with Initial System-Environment Correlations, Phys. Rev. Lett. \textbf{114}, 090402 (2015).

\bibitem{kos02} R. Kosloff and T. Feldmann, Discrete four-stroke quantum heat engine exploring the origin of friction, \textit{Phys. Rev. E} \textbf{65}, 055102(R) (2002).
 \bibitem{fel03} T. Feldmann and R. Kosloff, Quantum four-stroke heat engine: Thermodynamic observables in a model with intrinsic friction, \textit{Phys. Rev. E} \textbf{68}, 016101 (2003).
\bibitem{pla14} F. Plastina, A. Alecce, T. J. G. Apollaro, G. Falcone, G. Francica, F. Galve, N. Lo Gullo, and R. Zambrini, Irreversible Work and Inner Friction in Quantum 
Thermodynamic Processes, \textit{Phys. Rev. Lett.} \textbf{113}, 260601 (2014).


\bibitem{tor13} E. Torrontegui, S. Ibanez, S. Martinez-Garaot, M. Modugno,
A. del Campo, D. Guery-Odelin, A. Ruschhaupt, X. Chen, and J. G. Muga, Shortcuts to Adiabaticity, Adv. At. Mol. Opt. Phys. \textbf{62}, 117 (2013).
\bibitem{gue19} D. Guery-Odelin, A. Ruschhaupt, A. Kiely, E. Torrontegui, S. Martinez-Garaot, and J. G. Muga, Shortcuts to adiabaticity: Concepts, methods, and applications, Rev. Mod. Phys. \textbf{91}, 045001 (2019).

\bibitem{den13} J. Deng, Q.-H. Wang, Z. Liu, P. H\"anggi, and J. Gong, Boosting work characteristics and overall heat-engine performance via shortcuts to adiabaticity: Quantum and classical systems, Phys. Rev. E \textbf{88}, 062122 (2013).
\bibitem{cam14} A. del Campo, J. Goold, and M. Paternostro,  More bang for your buck: Super-adiabatic quantum engines, Sci. Rep. \textbf{4}, 6208 (2014).
\bibitem{bea16} M. Beau, J. Jaramillo, and A. del Campo,  Scaling-Up Quantum Heat Engines Efficiently via Shortcuts to Adiabaticity , Entropy \textbf{18}, 168 (2016).
\bibitem{aba17}  O. Abah and E. Lutz, Energy efficient quantum machines, EPL \textbf{118}, 40005 (2017). 
\bibitem{aba18} O. Abah and E. Lutz, Performance of shortcut-to-adiabaticity quantum engines, Phys. Rev. E \textbf{98}, 032121 (2018).
\bibitem{hou25} W. Hou, W. Yao, X. Zhao, K. Rehan, Y. Li, Y. Li, E. Lutz, Y. Lin, and J. Du, Combining energy efficiency and quantum advantage in cyclic machines, Nature Comm. \textbf{16}, 1 (2025).

\bibitem{com1} An alternative strategy would be to refill the 'entropic tank' of the engine by letting it thermalize with the external baths to the initial state $\rho (0) = \rho_\text{c} \otimes \rho_\text{h}$, before restarting it.
\bibitem{com2} This amounts to setting the relative entropy of coherence, $C_\text{r} (\rho) = S (\rho_\text{d}) - S (\rho)$, for system and baths to zero, since  $D ( \rho || \rho^\text{th} )= C_\text{r} (\rho) + D (   \rho_\text{d} || \rho^\text{th} )$ \cite{str17}.

\bibitem{str17} A. Streltsov, G. Adesso, and M. B. Plenio, Quantum coherence as a resource,  Rev.  Mod. Phys.  \textbf{89}, 041003 (2017).

\end{thebibliography}

\begin{thebibliography}{99}
\bibitem{poz18} A. Pozas-Kerstjens, E. G. Brown and K. V. Hovhannisyan, A quantum Otto engine with finite heat baths: energy, correlations, and degradation, New J. Phys. \textbf{20}, 043034 (2018).
\bibitem{ber17} M. N. Bera, A. Riera, M. Lewenstein, and A. Winter,  Generalized laws of thermodynamics in the presence of correlations, Nature Comm. \textbf{8},  2180 (2017).

\end{thebibliography}
\end{document}